# Quantum Field Theory and the Measurement Problem in Quantum Mechanics

Avi Levy[1], Meir Hemmo[2]

Draft

## Abstract

The measurement problem in quantum mechanics arises from the discrepancy between the unitary temporal evolution of quantum states described by the Schrödinger equation (in the non-relativistic case) and the non-unitary evolution ("collapse") of the quantum state in a measurement process. In quantum field theory, the standard interpretation of the S-matrix scattering formalism assumes that the in-state of an interacting system evolves unitarily to a superposition of all possible out-states and it "collapses" to a unique out-state only when a measurement is performed. Hence, quantum field theory too is subject to the measurement problem.

In this paper we propose a novel physical solution to the measurement problem based on quantum field theory. According to our proposal, in certain types of elementary interactions, in which the "particles content" of the system is changed (we explain this notion below) the temporal evolution is non-unitary. We argue that these interactions, which are almost instantaneous, lead to a genuine stochastic selection of an outcome subspace that has a distinct "particles content", but can be a superposition of momentum states, spin states, etc.

Our proposal is supported by Haag's theorem from which it follows that the existence of a unitary evolution from every free in-state to every free out-state of a non-trivial interaction is mathematically unsound. A version of Haag's theorem implies that a non-unitary evolution occurs in those processes where new types of particles are created and / or destroyed leading to a particles content change. Since in quantum mechanics such processes are excluded (because the particles content of a physical system is fixed), the appearance of a "collapse"[3] of the wavefunction in quantum mechanics seems mysterious. Not so, we argue, in quantum field theory.

---

[1] Philosophy Department, University of Haifa. levya@technion.ac.il
[2] Philosophy Department, University of Haifa. meir@research.haifa.ac.il
[3] When referring to our proposed non-unitary transition, we use the term "collapse" in square quotes to mark that it is radically different from the collapse of the wavefunction or the projection postulate in *standard* quantum mechanics, e.g., our "collapse" (as we shall see) is a stochastic process that occurs on extremely short time scales but it may not be instantaneous or abrupt.

We address and explain in detail the key concept of particles content change in quantum field theory (which requires clarification) as well as the locality properties of non-unitary processes. Finally, we show that in typical measurement processes, there is a well-defined non-unitary stage. Our proposal de-mystifies the projection postulate for measurement in the standard formulation of quantum mechanics in that it identifies the physical conditions under which the "collapse" occurs. We argue that non-unitary processes are not specific to "measurement" since they occur in other naturally originated processes in which there is a particles content change. We further argue that our proposal is consistent with all known experimental results predicted by quantum mechanics and quantum field theory.

**Keywords:** change of particles content; collapse of the quantum state in quantum mechanics; interpretation of quantum mechanics; interaction picture of quantum field theory; locality property of non-unitary processes in quantum field theory; measurement problem in quantum mechanics; scattering theory, renormalization and effective theories; unitary vs. non-unitary state transition.

## 1. Introduction

The measurement problem in standard quantum mechanics (von Neumann 1932) arises from the discrepancy between the unitary temporal evolution of quantum states described by the Schrödinger equation (in the non-relativistic case) and the non-unitary evolution ("collapse") of the quantum state caused by a measurement process. Since measurements are performed by physical processes and as such should also be described by the unitary Schrödinger equation, this discrepancy leads to internal inconsistency in the formulation of quantum mechanics.

In textbooks presenting quantum field theory, and even in the philosophical literature, the measurement problem is hardly mentioned. It may be due to a tacitly assumed "standard" interpretation of the S-matrix formalism in quantum field theory that stems from the corresponding scattering process in quantum mechanics from which it was derived. This interpretation suggests that the asymptotic[4] in-state of an interacting system evolves *unitarily* to a *superposition* of all possible asymptotic out-states and that it "collapses" onto a specific out-state only when a *measurement* is performed. This reading preserves the discrepancy in quantum field theory between the unitary and the non-unitary processes and hence the measurement problem persists.

---

[4] The meaning of asymptotic state is defined in Section 0. Roughly speaking, an asymptotic state is the state of the system long before or long after the interaction when the state is approximately free.

We suggest that this discrepancy can be accounted for by adding a "collapse" postulate that changes the standard interpretation of a *certain* type of interactions described by the S-matrix formulation in quantum field theory. Our proposal is very briefly this. We conjecture that there are two distinct types of interaction processes in quantum field theory. In one type, the state evolves "approximately unitarily"[5], but in the other type, in which (what we call) the *"particles content"* of the system is changed, the process is non-unitary. This latter case leads to a genuine *stochastic* selection of a subspace of the space of all possible outcome states. Admittedly, this notion of particles content needs clarification. As we proceed, we shall define it rigorously and give more details about the distinction between the above two types of processes in quantum field theory.

The crucial point of our proposal is that in the non-unitary processes, the asymptotic out-state of the interaction is *not* the superposition of all possible outcome states with non-zero amplitudes. According to our proposal, the out-state of an interaction in which there is a particles content change invariably has a determinate particles content (rather than superpositions of different particles content). Notice that the out-state may be a *superposition* of momentum states, spin states, etc. We suggest that the failure of the unitary evolution occurs in certain processes where according to quantum field theory new types of particles are created and others are destroyed. In the framework of quantum mechanics, such processes are excluded, simply because the particles content of a physical system is fixed by construction. But, in quantum field theory, where the fields' operators consist of combinations of creation and annihilation operators, a change in the content of particles is a standard phenomenon.

We suggest that in *every* measurement process such a non-unitary transition takes place, at a certain stage, and this leads to the so-called "collapse of the wavefunction" in standard quantum mechanics. In addition, we suggest that such "collapses" occur also in other physical scenarios in which there is a change of particles content, regardless of whether or not they are

---

[5] In quantum field theory there is no explicit description of the state evolution of an interacting system hence, it is unknown if there are exactly unitary state evolutions. However, there are certain types of interactions that has a unitary description in the non-relativistic limit in the framework of standard quantum mechanics. We use the term "approximately unitary" to differentiate these processes from the other type of processes which are proved to be non-unitary.

associated with measurements. Thus, the notion of measurement is not a primitive of our proposal.[6]

Giving up, even partially, the unitary temporal evolution of the states in quantum field theory seems like a painful renunciation of one of the pillars of the quantum mechanical framework. However, Haag's (1955) theorem indicates that the assumption that there exists a unitary transformation describing the temporal evolution from *every* free in-state to *every* free out-state of an interacting system is *unsound*. We take the theorem to support our proposal that in interactions, where particles are destroyed or created, the evolution of the quantum state is *non-unitary*. However, we argue that in the non-unitary scattering processes of quantum field theory, the calculation of the transition probabilities from an asymptotic in-state to an asymptotic ("collapsed") out-state is identical to those presented by the various methods of the S-matrix entries calculations. In quantum field theory there is no explicit temporal evolution description of these processes, and we do not purport to give one. This drawback is compensated by the fact that when the particles content of the interacting system changes, the duration of this stage of the interaction is extremely short and we take it that this fact explains why this stage cannot be directly observed. Hence, the empirical predictions of our interpretation are identical to the predictions of the standard reading of scattering processes in quantum field theory.

Yet, our interpretation has two advantages over the standard view of quantum field theory. Firstly, our proposal is *empirically falsifiable* since it is experimentally possible to detect superpositions of subspaces that correspond to different particles content, which our proposal rules out. In this sense our proposal leads to new experimental predictions by comparison to the stand conception of quantum field theory. Secondly, our proposal provides a much more coherent and physically based criteria for the occurrence of "collapses" in comparison to the ambiguous notion of "measurement" in the standard interpretation of quantum mechanics. Thirdly, our proposal leads to different experimental predictions than other collapse theories, e.g., the spontaneous localization theory proposed by Ghirardi, Rimini and Weber (1986; henceforth, GRW) in which external spontaneous "jumps" are added over and above the description given by the wavefunction.

---

[6] Other proposals relying on creation and annihilation of particles are e.g., (Danos and Kieu 1999; Diel 2015; Melkikh 2015; Maxwell 2018). However, these proposals ignore the key issue of non-unitarity in the state transition explained above (for more details about them, see Levy 2023).

In this paper, we address only the standard formulation of quantum field theory that uses the S-matrix formalism, including renormalization and effective theories. This theory has certain mathematical deficiencies; however, it does describe real phenomena and predicts with astonishing accuracy the experimental observations in particle physics. We do not consider algebraic quantum field theory (see Haag and Kastler 1964, Halvorson and Müger 2006), although it has a rigorous mathematical framework, since some of its assumptions do not hold in the interactions of the standard model (Haag 1996, Buchholt and Fredenhagen 2020).

The paper is structured as follows.

Section 2 presents the measurement problem and sets the stage for our proposal of resolving it within the framework of quantum field theory.

Sections 3 presents a very concise conceptual description of the scattering theory of quantum field theory, for those who are versed with the measurement problem in quantum mechanics but are not familiar with quantum field theory. We describe here free states and Fock spaces and presents the interaction picture in quantum field theory and the S-matrix formalism. It also discusses very briefly the issues of infinities that appear in computing the S-scattering matrix elements and how they are treated by regularization and renormalization methods.

Section 4 discusses issues concerning non-unitarity in quantum field theory. We present here the issue of nonequivalent representations, and of superselections rules and superselection sections as limitations on the superposition principle. We further describe the problematic consequences of Haag's theorem for the interaction picture, review certain attempts to restore unitarity in the scattering theory of quantum field theory, such as the Haag-Ruelle scattering theory, and discuss the impact of renormalization and effective theories on the non-unitarity issue.

Section 5 reviews the notion of "particle" in quantum field theory, beyond Wigner's concept of elementary particles as irreducible unitary representations of the Poincaré group. We discuss here extensions of the particle concept to improper particles such as infraparticles, compound particles (bound states), and present the notion of asymptotic (improper) particle content.

Section 6 proposes a definition for "particles content change". We argue here that the distinction we propose between approximately unitary and non-unitary processes in quantum field theory explains the discrepancy between the unitary evolution and the wavefunction "collapse" in standard quantum mechanics.

Section 7 discusses the two types of processes we propose in quantum field theory – the "approximately unitary" and the non-unitary (which occurs when the particles content of the system changes during the interaction). We give examples here for the non-unitary interactions in decay, absorption and emission and in scattering processes. Also, we explain the cluster decomposition principle and the *locality* property of these non-unitary interactions.

Section 8 discusses the local nature of our stochastic "collapse" postulate.

Section 9 argues that in every known measurement process, there is at least one non-unitary stage which causes the "collapse" of the wavefunction. We present here a scheme for a "standard" measurement where the "measured" eigenvalue of the (so-called) measured observable is actually being *calculated* from the detection of a certain particle in a localized space-time region (rather than being directly measured). We show in this section that our proposed scheme holds in a polarization / spin measurement. We give other examples in the appendix. The last part of this section points at the differences between our notion of "collapse" that is based on particles content change and the GRW spontaneous "collapse" theory.

Section 10 discusses three elementary phenomena that seem at first sight to challenge our proposal. We argue in outline why they do not in fact invalidate it.

Section 11 concludes the paper, and Section 12 contains two appendices: one on locality and the cluster decomposition principle and the other on non-unitary processes in measurement scenarios.

**2. The measurement problem in quantum mechanics and how it may be resolved by quantum field theory**

The measurement problem in quantum mechanics is manifested in an evident contradiction between the conflicting descriptions of two process types used by the standard (sometimes called Copenhagen) formulation of the theory:

**(i)** In type I processes, which are defined as "measurements" (or sometimes observations), the state of the system "collapses" on one of the eigenvectors (or eigenspaces) of the measured observable represented by a Hermitian operator over the Hilbert state space. The probability of observing a specific eigenvalue / eigenvector is given by the Born rule.

**(ii)** In type II processes the quantum state of a physical system is described by an element of a Hilbert space which evolves temporally in a unitary fashion e.g., according to the Schrödinger equation (in the non-relativistic case).

The problem of this formalism is that no clear boundary is drawn between the two types of processes, and the theory does not offer a physical way to distinguish between them (or to predict, or observe, when one type of process ends and the other begins). The contradiction lies in the fact that the stochastic collapse of the wave function dictated by type I processes cannot be described by the deterministic, linear and unitary time evolution that characterizes the type II processes. Hence, although a measurement (and an observation) is ultimately a physical process, it cannot be described by the standard formalism of quantum mechanics.

The measurement problem depends on two assumptions concerning the nature of physical processes which are not measurements:

1. The state space of a quantum system is a Hilbert space whose elements can be represented as superpositions of eigenvectors of every Hermitian operator (observable) on this Hilbert space.

2. The temporal evolution of these eigenvectors is given by a linear and unitary evolution operator defined by the Hamiltonian of the system.

In contrast to the unique Hilbert space description of quantum mechanics, there are interactions in quantum field theory (see the next section), which cannot be described in this way. As shown by Haag's theorem (Haag 1955), in various kinds of interactions, there is no one Hilbert space that contains both the initial state of the system and its final state, and there is no unitary evolution transformation that intertwines between these two states. One of the features associated with this result is that in quantum field theory there are annihilation and creation fields (operators) that can change the "particles content" of the system during an interaction.[7] Such a change is excluded in quantum mechanics where the system is defined by a fixed set of particles.

---

[7] The main reason for the non-unitarity is that in quantum field theory the state space has an infinite number of degrees of freedom, while in quantum mechanics the number of degrees of freedom is finite. The field structure of quantum field theory implies the existence of the creation and annihilation operators.

We propose that in every process which is considered a "measurement" in quantum mechanics there is a stage where this type of non-unitary interaction occurs. Although such interactions cannot be described by a unitary process in one Hilbert space, the probabilistic distribution of their outcomes can be calculated by the standard methods used in quantum field theory. They define an appropriate stochastic process between initial states and final states that is not unistochastic,[8] but can accommodate for what is perceived, in standard quantum mechanics as the "collapse of the quantum state".

**3 Free systems, scattering processes and interactions in quantum field theory**

Quantum field theory is a relativistic quantum theory and hence it combines the Hilbert state space formalism of quantum mechanics with the principles of special relativity. In this section we review very concisely the conceptual structure of relativistic scattering theory which is born from the merging of these two theories.

**3.1 Free particles systems in quantum field theory**

Combining the Hilbert space formalism with the requirement of relativistic invariance results in describing elementary systems in quantum field theory as irreducible unitary representations of the Poincaré transformations group. Wigner (1939) classified these representations (see also Tung 1985) and showed that they are characterized by two real parameters $m, \sigma$. The continuous parameter $m$ represents the rest-mass and the parameter $\sigma$ takes half-integer values and is related to the spin or chirality of the elementary system.[9] This classification is the basis for the concept of elementary particles in quantum field theory.[10] The Hilbert state space constructed for accommodating multi free particles systems is named Fock space which is defined as the tensor product $F(H) = \overset{\infty}{\underset{n=0}{\otimes}} H_n$, where the Hilbert spaces $H_n$ describe $n$ free particles state spaces. The elements of $H_n$ are denoted $\psi_\alpha$, where the index vector $\alpha = (p_1, \sigma_1, n_1, \ldots, p_n, \sigma_n, n_n)$ is defined by the momentum vector $p_i$ of particle $i$ (with $p_{i,0}^2 - p_{i,1}^2 - p_{i,2}^2 - p_{i,3}^2 = m_i^2$), the spin / chirality $\sigma_i$ of the $i$ particle and the parameter set $n_i$ which specifies the "type" of the particle. The empty state that has no particles is denoted $|0\rangle$ and is named the "vacuum".

---

[8] A unistochastic process has the property that its stochastic matrix entries decompose to $\Gamma_{i,j} = |U_{i,j}|^2$ where $U$ is a unitary matrix.
[9] Particles with integer spin / chirality values are named bosons and those with half integer values are named fermions.
[10] See Section 5 for a more detailed discussion of the particle concept in quantum field theory.

In a Fock space one can define *creation and annihilation operators* in the following way: The creation operator $a^\dagger(p,\sigma,n)$ adds a particle with properties $p,\sigma,n$ in the first location of the index vector $\alpha$, i.e. $a^\dagger(p,\sigma,n)|0\rangle = \psi_{p_1,\sigma_1,n_1}$, $a^\dagger(p,\sigma,n)\psi_\alpha = \psi_{[(p_1,\sigma_1,n_1),\alpha]}$. The annihilation operator is the adjoint of the creation operator and is denoted by $a(p,\sigma,n)$. It removes a particle with the properties $p,\sigma,n$ from states that include a particle with these properties, and it annihilates the vacuum $a(p,\sigma,n)|0\rangle = 0$.

Based on their definitions, it is possible to show that creation and annihilation operators satisfy the commutation relations (or anticommutation relations for fermions).

$$[a(\alpha'),a^\dagger(\alpha)] = \delta(\alpha'-q),\ [a(\alpha'),a(\alpha)] = 0,\ [a^\dagger(\alpha'),a^\dagger(\alpha)] = 0.$$

For each type of particle there is an operator valued field that describes its dynamics and interacting properties. These fields are created from a symmetric combination of creation and annihilation operators corresponding to the properties of this particle (see Duncan 2012 for a through presentation of this topic).

### 3.2 The interaction picture

Interactions are described in quantum field theory by the scattering formalism, originated in quantum mechanics, where an initial free in-state $\psi_\alpha^{in}$ evolves in time via a unitary process $U(\tau,\tau_0)$ to a final free out-state $\psi_\beta^{out}$. Both $\psi_\alpha^{in}$ and $\psi_\beta^{out}$ are members of the same Hilbert space and the subindices $\alpha,\beta$ contain the parameters defining these states as explained above. Roughly speaking, the vectors $\alpha,\beta$ characterize the "particles content" of the in-state and out-state respectively (see Section 6 for the details of this notion).

The in-states and out-states are assumed to be free states, since the interaction is well localized in space-time and its impact is negligible long before the interaction and long after it. The scattering formalism provides methods for computing transition amplitudes $S_{\beta\alpha} = \langle \psi_\beta^{out} | \psi_\alpha^{in} \rangle$ from an initial state with a given "particles content" to a specific final state

with another (or identical) "particles content". The matrix $S_{\alpha\beta}$ is a unitary matrix and the probability of transition from the state $\psi_\alpha$ to the state $\psi_\beta$ is given by $P_{\beta\alpha} = |S_{\beta\alpha}|^2$. In the Schrödinger picture the wavefunction describing the state of the system evolves according to $\psi(t,\mathbf{x}) = e^{-iHt}\psi(0,\mathbf{x})$, where $H$ is the Hamiltonian of the system. In the equivalent Heisenberg picture, the operators acting on the Hilbert space evolve according to $O(t,\mathbf{x}) = e^{iHt}O(0,\mathbf{x})e^{-iHt}$. The interaction picture combines the Schrödinger and the Heisenberg pictures and assumes that the full Hamiltonian can be decomposed into a free and an interaction part, respectively $H = H_0 + H_1$. Under mild conditions on the spatial behavior of the interaction potential it is possible to show that an interaction state (strongly) converges to a free state in the limits $t \to \pm\infty$ [11] (Duncan 2012, Sec. 4.3).

It follows that the unitary evolution operator describing the evolution of the interacting system from time $\tau_0$ to $\tau$ is given by $U(\tau,\tau_0) = e^{iH_0\tau}e^{-iH(\tau-\tau_0)}e^{-iH_0\tau_0}$, and hence $S_{\beta\alpha} = \langle \psi_\beta | U(+\infty,-\infty)\psi_\alpha \rangle$,.[12] As stated above, $|S_{\beta\alpha}|^2$ is the probability of transition from the normalized free in-state $\psi_\alpha^{in}$ to a normalized free out-state $\psi_\beta^{out}$. and from the unitarity of the evolution matrix $U(\tau,\tau_0)$, it follows that the total transition probability $\sum_\beta |S_{\beta\alpha}|^2$ is equal to 1.

The principle of relativity imposes two requirements on a theory of scattering. One is that its predictions must be similar for all inertial observers; the other is locality, i.e., an interaction localized in a certain region cannot have an impact faster than the speed of light on another space-like separated region. These two requirement restrict the form of the Hamiltonian that describes the interaction between certain particles. In essence, such a Hamiltonian is constructed as a polynomial in the fields of the particles participating in the interaction (see Duncan 2012).

---

[11] The free state is $\psi_f(x,t) = e^{-iH_0 t}\psi_f(x,0)$ and the interacting state is $\psi_I(x,t) = e^{-iHt}\psi_I(x,0)$, where at $t=0$ the Schrödinger and Heisenberg pictures coincide.

[12] The asymptotic states and the evolving states are related by:
$\psi_\alpha^{in} = U(0,-\infty)\psi_\alpha$, $\psi_\beta^{in} = U(0,\infty)\psi_\beta$.

By integrating the time evolution equation of the matrix $U(\tau, \tau_0)$, one can represent the scattering operator **S** defined by $S_{\beta\alpha} = \langle \psi_\beta | \mathbf{S} \psi_\alpha \rangle$ as the Dyson series

$$\mathbf{S} = I + \sum_{n=1}^{\infty} \frac{(-i)^n}{n!} \int_{-\infty}^{\infty} dt_1 \ldots dt_n T\{H_1(t_1, \mathbf{x}_1) \ldots H_1(t_n, \mathbf{x}_n)\},$$

where $T$ denotes the time ordering that reorders a product of operators in a decreasing time sequence (left to right). Using such asymptotic representations, the scattering matrix elements can be approximated by the Feynman rules or alternatively using discrete grid approximation procedures.

However, it turns out that there are serious difficulties concerning the computational aspects of the elements of the Dyson series since the integrals that define the terms of the series diverge. This is the notorious "infinities" problem that plugs the scattering theory in quantum field theory. We discuss this issue in the following section.

### 3.3 Infinities, regularization and renormalization

In this section we briefly describe some approaches that were developed in order to solve numerical issues that arise in the calculation of the transition amplitudes of the S-matrix. We present only conceptual aspects of these approaches that are relevant to our proposal for how quantum field theory may be taken to resolve the measurement problem in quantum mechanics.

Initially, these approaches were developed to treat non-convergence integrals that appear in perturbative based computations of the S-matrix amplitudes in the interaction picture. These infinite values originated from two types of mathematical singularities:

- Divergences of terms corresponding to very small distances, or equivalently very high energies (these are called *ultraviolet divergences*); and
- Divergences of terms corresponding to very large distances, or equivalently very law energies (these are called *infrared divergences*).

The methods developed for resolving these non-convergences include two steps: regularization and renormalization. Regularization is a mathematical procedure used to temporarily control the ultraviolet or infrared divergences. After regularization, it becomes possible to extract finite results from the perturbative series, order-by-order. However, any regularization technique introduces new arbitrary parameters (regularization scales). In the subsequent renormalization procedure these arbitrary dependences on regularization scales are absorbed by redefining the bare parameters of the theory (for a fuller explanation, see Collins 1984).

A "cut off" based regularization procedure may be implemented in the momentum domain by introducing an ultraviolet cut off as a finite value (< infinity), and infrared cutoff as an infinitesimal value (> zero). Thus, the elements of the perturbative series become functions of these cutoffs. In the renormalization stage the mass, fields and coupling constant of the Lagrangian are reparametrized by introducing counter-terms. The counter-terms are chosen, for each order, in such a way that those parts of the series' elements that diverge are removed. Then the cutoff value is taken to infinity, and this yields the final value for the computed perturbative expression at the computed order. Theories whose ultraviolet divergences can be systematically eliminated via a redefinition of a finite number of parameters in the Lagrangian are said to be *renormalizable*.

A more modern view is that these divergences provide evidence for the fact that quantum field theory is an effective theory in which physical quantities, such as mass and charge, depend on the scale of distances / energies relevant to the interaction described by the S-matrix. The existence of effective theories (represented by "effective Lagrangians") is remarkable as they succeed in screening off the effects of the dynamics of short distances. This may indicate that quantum field theory is an emergent theory, which approximates an underlying more fundamental theory with dynamical aspects at very short distances (high energies) that are beyond our current knowledge.

The "scale separation" techniques used in effective theories achieve adequate accuracy in describing low energy dynamics by smearing out the effects of large momentum components by "cut off" techniques. These techniques may be viewed as finite dimensional approximations, based on a finite number of "renormalized parameters" of an infinite dimensional theory. Alternatively, in non-perturbative approaches, an approximated theory may be defined on a discrete space-time lattice with finite extent that replaces the continuous infinite space-time and thus has built-in cut offs for the very high and very low energies.

It turns out that for "renormalizable" theories, the continuous limit of the finite dimensional approximation can be restored at the end of the amplitudes' calculations by: (i) Taking the ultraviolet cut off scale to infinity, or equivalently, taking the grid distance to zero without re-introducing the ultraviolet (and infrared) divergences; and (ii) Removing the infrared cut off by extending the finite spatial volume of the approximated problem to an infinite domain and adapting the normalization of the fields accordingly (see Duncan 2012, p. 362-3).

The long distance (in the spatial-temporal sense) behavior of Abelian gauge invariant theories, such as quantum electrodynamics, poses even more serious difficulties for computing transition amplitudes in quantum field theory. In fact, "For QED theory with a massive, charged particle (electron) coupled to exactly massless photons, the definition of a conventional Fock space and associated S-matrix fails in a fundamental way, the S-matrix vanishes identically in such a theory." (Duncan 2012, p. 713).

These infrared divergences and the vanishing of the S-matrix amplitudes arise because of mathematical idealizations of charged particles propagating in Minkowski space-time of infinite spatial volume and of detectors with infinitely precise resolution that can detect photons at infinitely low frequencies. Such divergences can be removed by introducing infrared cut off, i.e., lower limit to the photon energy, and thus modeling detectors with finite resolution (this type of resolution can be traced back to Bloch and Nordsieck 1937).

An alternative "grid based" method for computing finite values for the Dyson series terms is to compute the scattering amplitudes in a "box" of finite spatial volume. In this way the momentum integrals become discretized sums and thus introduce a natural infrared cut off of order of the inverse spatial size of the box. Inside the box the amplitudes are computed on a discrete grid with a parameter that specifies the distance between grid points. This parameter serves as an ultraviolet cut off since it limits the maximal interaction momentum values by a bound which is proportional to the inverse of the grid distance value. The values of the Dyson series terms become functions of the box's size parameter and the grid size parameter. It turns out that for renormalizable theories, when the first parameter is taken to infinity and the second parameter to zero, the values of the corresponding terms remain finite.

### 4 Non-unitarities in QFT Scattering Theory

Quantum mechanics is a theory with a finite number of degrees of freedom, and its formalism is based on a unitary evolution of the quantum (pure) states. By contrast, quantum field theory is a theory of fields that has an infinite number of degrees of freedom, and this leads to significant complications and to non-unitary processes. This non-unitarity is a key argument supporting our proposal for resolving the measurement problem. In this section we give a very concise account of issues related to non-unitarity in quantum field theory, focusing on the aspects that are relevant to our proposal.

## 4.1 Inequivalent unitary representations

In quantum mechanics a theorem by Stone (1930) and von Neumann (1931) asserts that under mild conditions all the unitary representations of a theory with a finite number of degrees of freedom obeying canonical commutation relations between position and momentum operators are unitarily equivalent. This mathematical fact indicates that the Schrödinger representation of quantum mechanics is unique in the sense that any other representation leads to the same physical predictions.

However, this uniqueness theorem does *not* extend to quantum field theory, which is a theory obtained not by quantizing a system of finitely many particles, but by quantization of fields, which are defined at every point of space(time). Even in the simple case of the Klein-Gordon field, non-equivalent unitary representations can be easily constructed. The problem presented by inequivalent unitary representations of the same physical theory is that they may have different physical content, and hence it is not clear which of them is the "correct" representation of the physical reality described by the theory.

It turns out that these inequivalent representations do have a physical meaning, for example in spontaneous symmetry breaking. In addition, it is impossible to describe non-trivial interactions without using inequivalent representations, since the representations of free and interacting systems are unavoidably unitarily *inequivalent*, as shown by Haag's theorem (see below).

## 4.2 Superselection rules and sectors

Superselection rules (SSR) were introduced by Wick, Wightman and Wigner (1952) (see also their 1970 for SSR for charge) as a limitation on the superposition principle of quantum mechanics and on the definition of observables.[13] A comprehensive review of SSR can be found in Earman (2008). Here we focus on elementary properties of SSR and their role in various attempts at solving the measurement problem (for SSR and the measurement problem; see Bub 1988).

---

[13] SSR did not play a central role in this theory. However, in the algebraic theory of quantum field theory, SSR became an essential part of Doplicher–Haag–Roberts reconstruction of quantum fields from the algebra of observables. For an overview of the DHR program, see Halvorson and Müger (2006) and references therein.

A simple SSR can be demonstrated by a Hilbert space H that decomposes as a direct sum of two orthogonal subspaces, $H \equiv H_1 \oplus H_2$, such that under the action of each observable the vectors in each $H_i$ are transformed into vectors only in the same $H_i$ for $i = 1, 2$, respectively. In other words, the action of observables in Hilbert space is reducible, which implies that $\langle \psi_1 | A | \psi_2 \rangle = 0$ for each $\psi_1 \in H_1$, $\psi_2 \in H_2$ and for all observables $A$. Relative to the given observables, coherent superpositions of states in $H_1$ with states in $H_2$ do *not* exist. A superposition of such $\psi_1$ and $\psi_2$ defines a mixed state rather than a pure state. SSR can be generalized to the finite or countable cases $H \equiv \oplus_i H_i$ and to the continuous case $H \equiv \int_\Lambda d\mu(\lambda) H(\lambda)$ (for more details, and for a connection to measurement; see Landsman 1995; for an algebraic version of SSR, see Giulinin 2009). The orthogonal subspaces $H_i$ are the *superselection sectors* of the physical Hilbert space $H$. As explained above linear combinations of states in distinct superselection sectors are forbidden and transitions between sectors are not only inhibited for the particular dynamical evolution generated by the Hamiltonian operator, but rather for *all* conceivable unitary dynamical evolutions due to conservation principles (see Duncan 2012, p. 85). Examples for such limitations are related to states of integer and half integer spin values (bosons and fermions), states of different electrical charge, and even states with mass differences, as indicated by the mass-shift version of Haag's theorem (see Section 4.3).

### 4.3 Haag's theorem

As we saw, quantum field theory has infinitely many degrees of freedom and it admits unitarily inequivalent Hilbert space representations. Haag's theorem follows from the fact that the vacuum state of the free and interacting theories belong to unitarily inequivalent spaces. The theorem renders the entire interaction picture mathematically *unsound* by implying that there cannot be a global unitary transformation connecting the states and field operators of a free and interacting theory, which the interaction picture is clearly predicated on.

A generalized version of Haag's theorem is the following (see Haag 1955; Hall and Wightman 1957; Jost 1961). If a scalar quantum field with non-trivial interaction and a positive mass, is unitarily equivalent to a free scalar quantum field, with a positive mass, then it is also a free field.

Since the unitary matrix $U(\tau,\tau_0)$ of the interaction picture transforms the incoming and outgoing free fields into the interacting fields, it constitutes a unitary transformation between the free and interacting spaces. It follows that those interacting fields are in fact free fields, and hence Haag's theorem seems to imply that the interaction picture breaks down.

Haag's result is not restricted to scalar fields. It is extended to Fermionic fields (Streater and Wightman 2000) and to the massless case (Pohlmeyer 1969). However, there are difficulties in extending it to gauge theories, since the Wightman axioms, on which it relies, do not hold for such theories (see Duncan 2012, p. 254). On the other hand, in gauge theories, like quantum electrodynamics and quantum chromodynamics, the interaction picture does *not* hold for other reasons, as explained in Section 4.5. It follows that despite its practical success, and due to its unitary nature, the interaction picture is not a sound mathematical picture of interacting systems in quantum field theory.

Haag's theorem implies that unitary processes cannot describe systems with free in-states and free out-states having non-trivial interactions. The apparent tension between this theorem and the phenomenal success of the interaction picture in explaining and predicting empirical observations calls for an explanation. There is a vast body of philosophical literature addressing this issue, which is comprehensively summarized in a recent review by Mitsch (2024). This review presents, analyses and compares a few approaches for relaxing this tension, suggesting that at least one of the assumptions on which Haag's theorem depends are not satisfied by scattering theories that are (or can be) used as an alternative to the interaction picture. For our proposed resolution to the measurement problem, the main issue is whether the unitarity of scattering processes can be restored, so we focus on this issue in the rest of this section.

Earman and Fraser (2006) propose to circumvent Haag's theorem by abandoning the assumption that the in- and out-spaces are free Fock spaces, and to adopt the Haag-Ruelle scattering theory instead (see Sec. 4.4 on this theory and on whether it resolves the non-unitarity issue).

Duncan (2012) and Miller (2018) argue that "Renormalization defuses Haag's theorem by breaking Poincaré invariance" (see Mitsch, Gilton and Freeborn 2024). However, Klaczynski (2016) argues that regularization and renormalization techniques describe non-unitary processes and suggests that assuming a unitary evolution in the scattering process should be abandoned. He argues that the process is not unitary by construction, since even in the simpler cases of renormalizable theories, with interaction Hamiltonian $H_I(x)$, the intertwiner map which has the form $V = \exp\left\{-iT\left(\int_0^t d^4 y H_I(y)\right)\right\}$ is not a simple exponentiation of the Hamiltonian due to the integration and the time ordering (see Klaczynski 2016 for a detailed analysis). Another more compelling argument raised by Klaczynski (2016) is based on a "mass shift" version of Haag's theorem (see Haag 1955; Duncan 2012, Sec. 10.5). This version states the following:

Let $\varphi_0, \varphi_1$ be two free fields of masses $m_0, m_1$, respectively, and assume that at a certain time $t$ there is a unitary map $V$, such that $\varphi_1(t,\mathbf{x}) = V\varphi_0(t,\mathbf{x})V^{-1}$. Then: $m_0 = m_1$, i.e., if $m_0 \neq m_1$, then there exists no such unitary map.

The renormalization and effective theories use effective Hamiltonians with new auxiliary interaction terms based on parameters that change according to the energy scale. One of the parameters is the rest mass of the fields, and hence these new terms introduce mass shift to the interaction and by the theorem quoted above there exists no unitary map between the in-state and the out-state[14]. Hence this "mass shift" version of Haag's theorem can be taken to imply that the mass parametrization in the effective theories' Hamiltonians excludes the possibility that a unitary process can describe these theories.

### 4.4 Haag-Ruelle scattering theory

The interaction picture supports a coherent asymptotic definition of particles based on postulating that the in- and out-spaces are identical Fock spaces. However, Haag's theorem implies that due to unitarity this postulate is incompatible with the possibility of describing non-trivial interactions. The Haag-Ruelle theory is aimed at restoring the essential concept of a unitary evolution by compromising on the Fock space postulate and replacing the definition of free particles states with *asymptotic* particles states.

Haag (1958) and Ruelle (1962) propose a scattering theory that establishes the existence of well-defined unitary evolution from the in-states to the out-states of an interacting field. However, these in-states and out-states are not "free" (i.e., they do not belong to a Fock space) but are asymptotic states in the interaction Hilbert space $\mathrm{H}$. Haag and Ruelle constructively show that there are in- and out-subspaces $\mathrm{H}^{in}, \mathrm{H}^{out} \subset \mathrm{H}$ which are isomorphic to dense subspaces of appropriate Fock spaces $H_F^{in}, H_F^{out}$. They construct a smeared version of the interaction field that produces a time-independent asymptotic one-particle state vector in H with definite momentum properties (see Duncan 2012, Sec. 9.3 for the details of this construction).[15]

---

[14] Klaczynski (2016, Sec. 17) claims that renormalization circumvents Haag's theorem, since it assumes that there exists a unitary intertwiner between the free and interacting fields. Our view on Haag's theorem is that it excludes the existence of a unitary map between the in-state and the out-state which is just what is proved in this section.

The Haag-Ruelle scattering theory is based on the same Wightman axioms that are used in the proof of Haag's theorem with the additional assumption that the interacting fields' masses are positive. The theory was extended by Buchholz (1975, 1977) to include massless particles of spin 0 and $1/2$, but the case of spin 1 gauge bosons was not resolved.

It is important to note that the Haag-Ruell scattering theory is constructed in such a way that *there i*s a unitary evolution from the asymptotic in-state to the asymptotic out-states. However, this theory cannot describe the gauge invariant interactions of the standard model and is not used in the actual calculations of the amplitudes that are measured in real experiments. Therefore, it is of theoretical value, but probably does not represent real world scattering processes, which according to our proposal, are not unitary, whenever the interactions induce a change of particles content (see additional details in the next sections).

**4.5 Non-unitarity beyond Haag's theorem**

As explained above both Haag's theorem and the Haag-Ruelle theory that attempts to amend the theorem's impact on the interaction picture do not hold in the case of gauge invariant theories. Since the three fundamental forces in the standard model are described by theories of this type, they deserve a separate discussion. We focus here on quantum electrodynamics that describes electro-magnetic interactions.

It turns out that for positive mass particles with non-localizable electro-magnetic charge, such as the electron and the proton, Wigner's concept of a particle does *not* hold, since electrons and protons they are inevitably accompanied by infinite clouds of massless photons. Such particles are named *infraparticles* (see Schroer 1963) and it was shown by Buchholz (1986) that, as a consequence of Gauss' law, pure states with an abelian gauge charge can neither have a sharp mass nor carry a unitary representation of the Lorentz group. In fact, single-particle electron or proton states with differing momentum fall into unitarily *inequivalent* superselection sectors. In particular, the matrix elements of all local operators of the theory (and not just of the S-operator, as discussed above) vanish between such states. This unitary inequivalence is far more consequential than the one implied by Haag's theorem, since it is impossible to construct well-defined normalizable single-electron wave-packets as different momenta states of the electron lie in inequivalent sectors (see Duncan 2012, p. 722). The case of non-Abelian gauge theories is even more subtle, and we will not address it here; see Buchholtz (1996) for a basic discussion of these cases.

---

[15] The Haag-Ruelle formalism can be extended to incorporate theories of bound states.

We conclude this section by noting that the evidence so far presented here strongly indicates that the interaction picture (including the regularized and renormalized versions) describing the scattering processes of the standard model *cannot* be based on a unitary evolution. This is true for every non-trivial interaction and especially for the gauge invariant ones. The approaches attempting to restore the unitarity of scattering processes are inapplicable to the way in which scattering amplitudes are computed in practice and they fall apart completely for gauge invariant theories.

**5 The concept of a particle in quantum field theory**

All approaches for formulating scattering scenarios have in common the paradigm that there are asymptotic in- and out-states that can be detected as "particles" long times before or long times after the interaction. However, they allow for a broader concept of a particle than just those states created by the operation of a quantum field on the vacuum. The extended notion of particle includes bound states (see Zimmermann 1958), e.g., a Hydrogen atom or an ion of a stable atom, and also infraparticles as described in the previous section. It is therefore clear that Wigner's concept of a particle as an irreducible representation of the Poincaré transformation group is too narrow for covering all asymptotic particle-like states appearing in quantum field theory. Moreover, the notion of "particle", represented by either a free or an asymptotic state which is used by the scattering formalism is not well-defined. Since our proposal for resolving the measurement problem relies on the concept of particles content change, we investigate the notion of a particle more rigorously in this section.

A generalization of Haag-Ruelle theory establishes the existence of scattering theories of massive particles (in the Wigner sense) with localized charges (see Dybalski 2005). However, for massive particles with a non-localized charge, there are *no* local operators interpolating between the vacuum and the single-particle state. To overcome this difficulty, Bloch and Nordsieck (1937) proposed to model macroscopic particle detectors as a positive and "almost local" operators that do not respond to the vacuum. Araki and Haag (Araki 1999) show that a theory can be constructed where scattering states have indeed the desired "particle" interpretation with regard to the observables representing macroscopic particle detectors (see Buchholtz 2023 for additional details). Buchholz (1977) proposed an extension for massless particles, such as the photon, of a resembling scattering theory.

However, such scattering theories cannot be extended to particles carrying an Abelian gauge charge, such as the electron and the proton, which are inevitably accompanied by infinite clouds of ("on shell") massless particles (i.e., infraparticles). Buchholtz (1986) showed that the states representing these infraparticles cannot be described by a unitary representation of the Lorentz group, essentially because in these theories, single-particle electron states with differing momentum fall into unitarily inequivalent spaces. Therefore, well-defined normalizable single-electron wave-packets states cannot be constructed, and there is no one-electron state corresponding to the squared electron mass. For these infraparticles the usual Haag–Ruelle scattering theory, which depends heavily on the existence of a mass gap and normalizable single-particle states, becomes inapplicable.

The difficulties in defining a sound asymptotic particle concept are not limited to abelian long-range interactions and they appear also in non-Abelian gauge theories at very small spacetime scales due to the confining forces. The models for describing these cases require a quite different treatment than those described above and will not be discussed here (see Buchholtz 1996 for additional details).

Buchholtz (2023; see also references therein) describes attempts at defining a scattering theory for "improper particles" based on improper particle states of sharp energy-momentum $p$ that can be defined by considering a special type of localizing operators. However, a general scattering theory based on improper particle states has not yet been developed.

Other important and intricate issues concerning the particle concept in quantum field theory are the distinction between stable and unstable particles and the distinction between elementary and composite particles (see Duncan 1996, Sec. 9.6). For our purposes it is enough to note that, depending on the specific context and scale of the scattering process, all combinations of stable / unstable and elementary / composite particles may appear as the asymptotic states of the interaction. For example, it may be sensible to view an unstable particle, with an average lifetime much larger than the other timescales of interest in the scattering processes under study as a stable particle and include it in the asymptotic states of the theory.[16]

The elementary / composite distinction depends on the way the Hamiltonian of the interaction is written. If the exact interacting dynamics, without approximation, has a precise *finite* expression in terms of products of local fields at a single spacetime point, then any particle interpolated by such a composite field can be considered elementary. For example, the ground state of the composite hydrogen atom and the proton which is composed of three quarks may be considered as out-states representing a detectable particle. "From the point of view of the asymptotic formalism of field theory, a stable composite particle is on just the same footing, in being present in the in- and out-spaces of the theory, as a stable elementary particle of the theory." (Duncan 1996, p. 299).

To conclude this section, we mention that a general scattering theory based on stable improper asymptotic particle states (elementary or composite) has not been developed yet. However, it is possible to formulate a weaker concept of (improper) "particles content completeness" which states that the measured and conserved quantities, such as energy-momentum, charge and spin, can be extracted from the asymptotic particles content of the system (Buchholz 2023). Despite the fact that even such a weaker concept has not yet been fully rigorously defined, we will adopt it in the rest of this paper (as a working hypothesis; see our formal definition below). The justification for using this notion is that in practice the detection of "particles" is explicitly or implicitly assumed by the standard interpretation of measurement outcomes of every scattering experiment.

### 6. Particles content change and non-unitary processes

As we shall stress below, we take Haag's theorem and the other non-unitary issues discussed above to imply specifically that the description of interactions in which there is a change of particles content cannot be unitary. It is precisely here that we shall apply our suggested new "collapse" postulate of a *non-unitary stochastic selection* of an outcome state with a distinct particles content, with probabilities given by the S-matrix formalism.

Before presenting the formal definition of particles content change it is important to mention again that the concept of an improper particle includes not only elementary particles and improper elementary particles, such as an electron with a cloud of photons, but also bound states that are composed of a few elementary particles. Such cases can be considered as single particles in the context of certain interactions.[17]

---

[16] For example, the neutron mean lifetime of 15 minutes can be considered *infinite* in comparison to most subatomic processes.

In what follows we adopt the notation presented in Section 3 for the Fock space free states for describing the (improper) particles content of a scattering process asymptotic states based on the assumption of "particles content completeness", as explained in Section 5. This means that such particles are *defined* by a finite set of concrete values of momentum and spin and a set of parameters (e.g., charges).[18]

Denote the $k$-particles in-state by $\psi^i_{p_1,\sigma_1,n_1...p_k,\sigma_k,n_k}$ and the $l$-particles out-state by $\psi^o_{q_1,s_1,m_1...q_l,s_l,m_l}$. The subindices $p_i, q_j$ denote the momentum; $\sigma_i, s_j$ denote the spin / chirality; and $n_i, m_j$ denote parameters that define the specie of particles in the in-state and the out-state respectively, where $i = 1,\ldots,k$, $j = 1,\ldots,l$.

Definition: (improper) particles content change:

If in the interacting system defined above there is at least one in-index $i$ for which there is no out-index $j$, with $n_i = m_j$, or at least one out-index $j'$ for which there is no in-index $i'$, such that $m_{j'} = n_{i'}$, then the interaction changes the particles content of the system.[19]

From now on, when we say "particles content change" we refer to the broader sense defined above; also, for brevity we shall omit the prefix "improper". Certain challenges following this definition will be discussed in Section 0.

## 7. Approximately unitary and non-unitary processes in quantum field theory

A scattering interaction, defined by a certain Hamiltonian or Lagrangian, has invariably a positive probability for resulting in an out-state with identical particles content to the in-state. Since in this case the particles content is fixed, there is a common Hilbert space, where both the in-state and the out-state are its elements. In the non-relativistic case, there is a unitary transformation that intertwines them. Such unitary processes have temporal evolution operator $U(\tau_0, \tau)$ that describes the evolution of the system for relatively long periods of time. For example, in the non-relativistic description of the Stern Gerlach experiment, the particle's trajectory through the magnetic field can be described by the Schrödinger equation. In quantum field theory, in relativistic high energy cases, there is no explicit description of the temporal evolution of the state even if the particles content does *not* change, since the S-matrix formalism allows for computing transition probabilities, but does not describe the state evolution. However, we assume, on the basis of the non-relativistic theory, that in this case there is *no* "collapse" of the state, even though quantum field theory does not provide an explicit general mathematical expression for the temporal evolution. Since we are interested in addressing the measurement problem in non-relativistic quantum mechanics, we can avoid delving into the (unknown) details of the relativistic temporal descriptions of these types of interactions.

---

[17] The proton and the neutron are composed of three quarks. A hydrogen atom is composed of proton and electron, etc.
[18] The definition for a superposition of such states is a straightforward generalization of this definition.
[19] Note that the number of identical particles may change between the in-state and the out-state *without* affecting the particles content of the system.

In the other cases,[20] where the particles content of the out-state is different from that of the in-state, the discussion in the previous sections shows that there *cannot* be a unitary transformation that maps the in-state to the out-state. This seems to violate one of the foundations of quantum mechanics and quantum field theory, since the unitary formalism leads to the probabilistic structure of these theories. However, what is required for a scattering process to have a consistent probabilistic structure is only that the sum of all the probabilities of the interaction outcomes accumulates to 1. Such interactions *can* be described by a stochastic matrix $\Gamma$, where the probability that the out-state of the interaction is $\beta$, given that the in-state is $\alpha$, is given by $\Gamma_{\alpha\beta}$. However, as explained in Section 4, Haag's theorem implies that for any non-trivial interaction this matrix is not generated by a unitary evolution that intertwines between the in- and out-states.

Also, in contrast to the non-relativistic unitary case, where physical processes may have relatively long durations, the time span of the non-unitary processes, where particles content is changed, is *extremely* short. Depending on the type of interaction, it is of the order of $10^{-10}s$ for the weak interaction, $10^{-16}s$ for the electromagnetic interaction and $10^{-23}s$ for the strong interaction. Therefore, according to our proposal, the unitary paradigm of quantum mechanics fails for almost instantaneous processes *only*, where the evolution of the state cannot be practically traced.

---

[20] Note that a scattering process which is defined by a certain Hamiltonian may have an out-state with particles content identical to that of the in-state, but it may also have *other* out-states, where the particles content of the system is changed.

Once all these details are in place, here is our proposal: the non-unitary and the approximately unitary types of processes in quantum field theory correspond to the von Neumann type I and type II processes in quantum mechanics, respectively. We propose that the "collapse" of the state in type I processes corresponds to a stochastic "selection postulate" which is manifested in the non-unitary process of quantum field theory. According to this postulate, when a particles content change occurs, the out-state of the non-unitary process of an interaction is *not* the superposition of all possible out-states with different particles content but is a state with only *one* distinct particles content. The reason is that states with different particles content belong to different *inequivalent* Hilbert subspaces (superselection sectors) and hence *cannot* be superposed. Notice that these out-states are in general superpositions of states with identical particles content over momentum, spin and even particle numbers (for bosons).

In this way von Neumann's "collapse" postulate in a measurement process of standard quantum mechanics is grounded, according to our proposal, in the fact that in every process that may be considered a measurement, there is at least one stage in which a non-unitary interaction with particles content change occurs.

### 7.1 Non-unitarity in processes of decay, absorption and emission, and scattering

In this section we analyze and demonstrate the non-unitarity of three common types of interactions described by quantum field theory in processes of decay, absorption and emission, and scattering.

### 7.1.1 Particles content change in a decay process

Consider a decay of a particle in an unstable state. Denote by $|P_u\rangle$ the in-state long before the interaction. In every time interval there is a positive probability that a decay does *not* occur, and the unstable state persists, and a complementary probability that a decay occurs and the particle decomposes into a stable particle $P_s$ and a sub-particle $Q$. In this second case, the state is denoted $|P_s\rangle \otimes |Q\rangle$.

A unitary evolution description of this process, based on quantum mechanics, may be given by: $e^{-\tau t}|P_u\rangle + \sqrt{1-e^{-2\tau t}}|P_s\rangle \otimes |Q\rangle$, where $\tau$ is the decay lifetime parameter.

A "collapse" version of this process is given by $|P_u\rangle \to |P_s\rangle \otimes |Q\rangle$. The symbol $\to$ means

$$\begin{cases} |P_u\rangle & t \leq t_d \\ |P_s\rangle \otimes |Q\rangle & t_d + \tau_I < t \end{cases}$$, where $t_d$ is an exponential random variable with parameter $2\tau$ and $\tau_I \ll \tau$ is the time interval in which the decay actually occurs.

The unitary description has two faults. First, the probabilities of detecting the unstable particle state or the decayed particle state in a time interval of a certain length are *not* invariant under time translation. Second, the transition *cannot* be described by a unitary evolution operator of the form $e^{-iHt}$, where $H$ is a Hermitian Hamiltonian, since such descriptions cannot produce a temporal exponential decay of the state's coefficients.[21] Therefore, a standard unitary evolution cannot rigorously describe a simple decay process.

By contrast, the "collapse" based description is consistent with quantum field theory, where the decay rate is derived from the S-matrix formalism, which "… makes sense only if the time $\tau_I$, during which the interaction acts, is much less than the mean lifetime ($\tau$) of the (in) particle." (Weinberg 1996, p. 137). This condition asserts our claim that the decay interaction occurs in a very short time. The almost instantaneous interaction is a *non*-unitary process that leads to a particles content change, from an in-state of an unstable particle to an out-state of a stable particle and an emitted sub-particle.

### 7.1.2 Particles content change in photon absorption and emission processes

In an absorption process, where an atom interacts with a photon, the atom is exited or else an electron is emitted from the atom. If the atom immediately decays into its ground state by emitting a photon, then the asymptotic out-state has the same particles content (an atom and a photon) as the in-state. However, if an electron $e$ is emitted, then the in-state includes an atom $A$ and a photon $\gamma$ and the out-state is composed of an ion $A^+$ and the emitted electron, in which case a particles content change has occurred. The "collapse" description iof this case is: $|\gamma\rangle|A\rangle \to |e\rangle|A^+\rangle$, where $\to$ has the same meaning as in the previous subsection. A unitary evolution description must assume that the in- and out-states are superposed $a(t)|\gamma\rangle|A\rangle + b(t)|e\rangle|A^+\rangle$, where $a^2(t) + b^2(t) = 1$, $a(t) \underset{t \to -\infty}{\to} 1, b(t) \underset{t \to -\infty}{\to} 0, a(t) \underset{t \to \infty}{\to} 0, b(t) \underset{t \to \infty}{\to} 1$.

---

[21] A common *ad hoc* "solution" is to add a non-Hermitian part to the Hamiltonian, but such a Hamiltonian does not conform to the principles of quantum mechanics.

However, such a description indicates that the atom and its ion exist in a superposition state even before the photon hit the atom which is an absurd.

Similarly, when an energetic particle hits an atom and emits a photon, the interaction results in a particles content change. In this case, the in-state includes the atom and the hitting particle, and the out-state includes the atom, the hitting particle and a photon $|P\rangle|A\rangle \rightarrow |P\rangle|A\rangle|\gamma\rangle$.[22] The unitary description is given by the Weisskopf-Wigner approximation of the decay of an atom to its ground state, resulting in a photon emission. A simplified version is given by $a(t)e^{-i\omega_0 t}|P\rangle|A\rangle|0\rangle + b(t)e^{-i\omega_p t}|P\rangle|A\rangle|\gamma\rangle$, where $|0\rangle$ is the vacuum state which is introduced in order to explain the creation of the photon. This description suffers from the same faults described in the previous subsection.[23]

### 7.1.3 Particles content change in scattering process

Let us look at a simple scattering interaction. Assume an idealized interaction with an in-state $|in\rangle$ and three possible out-states: $|in\rangle$ with probability $P_0$; $|out_1\rangle$ that has a certain particles content that occurs with probability $P_1$; and $|out_2\rangle$, with a different particles content with probability $P_2$, where $P_0 + P_1 + P_2 = 1$. As a concrete example, consider an energetic photon $\gamma$ that creates (in the vicinity of a heavy atom) a pair of electron-positron $|e\rangle|e^+\rangle$ or a pair of muon and anti-muon $|\mu\rangle|\mu^c\rangle$. The collapse description of this transition is *either* $|\gamma\rangle \rightarrow |\gamma\rangle$ or $|\gamma\rangle \rightarrow |e\rangle|e^+\rangle$ or $|\gamma\rangle \rightarrow |\mu\rangle|\mu^c\rangle$ with probabilities $P_\gamma$, $P_e$ and $P_\mu$, respectively. The unitary evolution description should be of the form $a(t)|\gamma\rangle + b(t)|e\rangle|e^+\rangle + c(t)|\mu\rangle|\mu^c\rangle$, with the obvious restrictions on $a(t), b(t), c(t)$.

The "collapse" description is consistent with the S-matrix formalism of quantum field theory. However, there is no formulation in quantum field theory that conforms with the unitary evolution description as such a description negates Haag's theorem.[24]

---

[22] There is an intermediate state where an electron has jumped to a higher level by the impact of the impinging particle and the atom was exited. But since the emission of the photon occurs almost instantaneously, we ignore this stage.
[23] This description does not conform with standard quantum mechanics theory where a vacuum state has no meaning
[24] In quantum mechanics creation and annihilation of particles cannot be formulated at all.

The essential difference between these two descriptions is that in the standard unitary picture the superposition out-state "collapses" only when a "measurement" is performed, but in the non-unitary description the collapse occurs in the almost instantaneous moment when the particles content changes.

These examples provide scenarios where our "collapse" interpretation can be verified experimentally. In order to disprove the validity of the non-unitary description it is required that a superposition of two states with different particles content is experimentally detected. While states of superposition between spin and position have been detected, as far as we know, a *superposition state of different particles contents* (according to our definition) has *not* been observed. There are certain cases that seem to challenge this assertion. We will discuss them in Section 10 and argue that given the current state of art they are *not* counter examples to our approach.

## 8. The locality property of non-unitary processes

A unitary evolution process can describe non-localized effects, in both space and time, as in the Stern-Gerlach experiment, where the wave-function of the particle splits to two non-overlapping packets by the effect of the magnetic field. However, the "collapse" induced by the non-unitary processes is a *local* effect, i.e., immediately after its occurrence the outcome state particles are well localized in space-time, and this calls for an explanation.

As explained in the previous sections, the temporal duration of the non-unitary processes is extremely short, which makes them temporally localized. In order to validate the space-time local nature of the "collapse" it is required to show that the "collapse" is also localized in position. This localization is partially proved by the cluster decomposition principle of quantum field theory, according to which interactions carried out in space-like separation cannot influence each other and hence the interactions are localized. However, the proof of the cluster decomposition principle assumes that the in-states are localized. Hence, it does not explain why, for example, in the double slit experiment the particle which has a spread wavefunction, nevertheless interacts with the screen at a single position only.

Therefore, a more subtle analysis is required for cases where, for example, a particle is described by a quantum mechanical wavefunction that has two picks at a certain time, with equal absolute amplitudes, located at positions far apart from each other. At these two positions two detectors are located, one in each position, and *prima facia* it is possible that the particle interacts, non-locally, with *both* detectors. In the appendix (Section 12.1) we present an analysis based on the cluster decomposition principle and momentum conservation which shows that such a non-local interaction is *impossible*.

**9. What is being measured in a measurement process?**

In this section we argue that in every measurement process there is at least *one* stage where a non-unitary interaction, which changes the particles content, occurs. As explained in Section 0, this change, which is localized in space-time, has the appearance of a wave-function "collapse" in the quantum mechanical description of the measurement process. In our approach these non-unitary interactions are not restricted to measurement processes (*per se*) but occur in natural processes, as demonstrated in Section 7. Therefore, the postulate that we suggest (in Section 7) for the occurrence of a stochastic "collapse" originates in concrete and well-defined physical conditions, and is not imposed by a non-physical vague concept of a "measurement".

The non-unitary stage of a measurement is characterized, as explained in Section 8, by its locality and by the occurrence of particles content change.[25] This allows for particles detectors to detect the existence of certain particles in a (usually well localized) space-time region. We suggest that in general what is actually being measured in a measurement process is the *presence* of a certain particle in a certain space-time region. Therefore, we will demonstrate, later in this section, that the observable eigenvalue which is "measured", e.g., spin, momentum, etc. is actually being calculated or inferred from the details of the details of the interactions and the information of *detecting a specific particle in a certain space-time region*.[26] There are additional two features of measurements, which guarantee that a recordable macroscopic outcome is attainable:

1. The non-unitary interaction has only a negligible probability to reverse in a very short time; otherwise, the outcome may become undetectable (e.g., when an electron and positron pair is created, but almost instantaneously it is annihilated).

---

[25] It is crucial to note that in our approach, despite the locality aspect, the "collapse" is *not* on the position basis, but on a Fock-like basis of asymptotic free states. For more details see the analysis in this section of the difference between our approach and the GRW approach.

[26] However, by this we do not mean to say that the non-unitary "collapse" occurs necessarily in this detection. In our approach, the "collapse" is brought about as a result of a change of particles content that may occur before and / or during the detection of particles.

2. The final, detectable out-state is stable on a macroscopic time scale, i.e., it is not composed of unstable particles or resonances that immediately decay. When such cases occur only the final stable outcomes are detected, and then given the details of the interactions, intermediate stages or the existence of resonances may be deduced from them.

Since there is no rigorous definition in quantum mechanics of what a measurement process is, it is difficult to prove that in every measurement there is a non-unitary stage. Therefore, it is only possible to propose a general scheme for a measurement scenario and then check whether or not the scheme holds in a variety of measurement scenarios. The scheme proposed below assumes that a certain observable, represented by the eigenvalues of a Hermitian operator, is being measured. The general scheme we propose of a measurement process[27] includes the following stages:

1. The system (e.g., a particle) is manipulated in such a way that the eigenvalues of the measured observable are coupled to distinct wavefunction components corresponding to non-overlapping regions of space-time.

2. A non-unitary interaction is performed as part of a measurement procedure in which a particle(s) is detected in one (and only one) of these nonoverlapping regions. The local nature of the non-unitary interactions ensures that the particle is detected in one region only.

3. The detection signal is amplified and recorded, and the result of the measurement is set to be the eigenvalue corresponding to the region where the particle was detected.

A simple example is the measurement of light polarization or of a particle's spin:

1. In the first stage the photon/particle beam is split into two packets each corresponding to a unique polarization/spin value.

2. Then, in the second stage, one (and only one) of two photon/particle detectors, located respectively at the two packets' regions, detects the photon/particle. The detection is accomplished by a non-unitary absorption of the photon or by a non-unitary interaction in which the particle ionizes an atom and an electron is released. In this stage the "collapse" on the Fock-like basis is accomplished since in the activated detector an interaction with particles content change occurs. Now, from this detection the location of the particle is known and by the coupling between spin (or polarization) and location the spin/polarization value can be deduced.

---

[27] An actual measurement procedure may include a few processes of this type and the accumulated information from all these processes is used to calculate the eigenvalue of the measured observable.

3. The signal from the activated detector is amplified, and the polarization/spin value corresponding to its location is recorded.

In the appendix (Section 12) other measurement scenarios are analyzed and the stage in which the non-unitary interaction occurs is indicated.

We conclude this section by pointing out that our collapse proposal based on quantum field theory may seem to resemble the GRW (See Ghirardi, Rimini, and Weber 1986) theory of spontaneous localization ("collapse" onto the position basis). However, there are three major differences between the two approaches:

(i) The "trigger" for the collapse in the GRW theory is a global external stochastic mechanism, added over and above the facts described by standard quantum mechanics. In the GRW theory there are two additional *new* constants of nature: the probability for the occurrence of a "collapse" at a given time (see below) and the width of the GRW Gaussian at which the localization is likely to occur. These two facts are assumed over and above the facts described by quantum mechanics. By contrast, in our proposal, the "collapse trigger" is based on the description of elementary interactions in contemporary quantum field theory. It is a consequence of Haag's theorem that non-trivial interactions in quantum field theory *cannot* be unitary, and so all we do is identify the non-unitary stochastic element in the current best theory of the elementary interactions. We do *not* introduce additional physical facts but rather fill in some missing gaps in the present theory which turn out to resolve the measurement problem in quantum mechanics. In this sense, our proposal is a sort of a bottom-up recovery of standard quantum mechanics by the nature of elementary interactions (i.e., those interactions resulting in a change of particles content) that are already known in quantum field theory.

(ii) In the GRW theory the collapse is on a narrowly peaked Gaussian in the *position basis*, while in our proposal the preferred basis for the collapsed states is a *Fock-like basis* of asymptotic free states. Indeed, due to the local nature of the non-unitary processes the two approaches seem similar, since the "collapse" on the Fock-like basis occurs in a well-localized space-time region. However, in quantum field theory a position basis does not even exist (see Duncan 2012, Sec. 6.5), hence a collapse on the position basis in the interactions described by the theory is meaningless.

(iii) In the GRW theory the probability for the occurrence of a "collapse" at a given time and region of space is proportional to the number of particles occupying this region at the designated time. This means that the probability for the occurrence of a "collapse" becomes significant only when the interaction involves a many-particles system (e.g., a standard measuring device). By contrast, in our proposal the "collapse trigger" has significant probability according to quantum field theory itself in common *elementary* interactions.

## 10. Challenges for the "collapse" postulate in quantum field theory

Our "collapse" postulate states that the out-state of a non-unitary interaction in quantum field theory has a distinct particles content and is not the superposition of all possible out-states (with non-zero amplitude). Let us examine now three challenges that seem to require further investigation.

### 10.1 Superposition of states with different photon numbers

The first challenge is posed by states of photons that are superpositions of distinct eigenstates of the number operator.[28] However, our definition of particles content change does not include the formation of such states. The condition it requires is that at least one new type of particle is created and one type annihilated. But an interaction that creates a superposition of states with different number of bosons[29] does not satisfy this condition. Therefore, it does not trigger a "collapse" and hence does not contradict our "collapse" postulate. Also, these superposition states of photon number are unstable and require maintaining by an external macroscopic control. Therefore, they cannot be considered as stable asymptotic free states of an elementary interactions, as we require.

### 10.2 Particle oscillations

The second challenge to our "collapse" postulate is posed by the experimental phenomena of particle oscillations. These phenomena can be classified into two types: particle-antiparticle oscillation (e.g., $K^0 \leftrightarrow \bar{K}^0$) and flavor oscillation (e.g., neutrino oscillation). Since the out-state of certain interactions includes such oscillations between *different* types of particles it seems to contradict our claim that the out-states of non-unitary interactions are *not* superpositions of states with distinct particles content.

---

[28] The existence of such states was reported in the literature in recent years in the framework of quantum optics (see Loredo 2019; Stammer 2022; Rivara-Dean 2022).
[29] As far as we know, there is no evidence for such states for fermions.

Two aspects of these phenomena resolve this apparent contradiction. Firstly, these oscillations are described in an *ad-hoc* framework based on a unitary matrix (PNMS matrix for neutrinos, CKM matrix for quarks, and another $2 \times 2$ matrix for the Kaons) that describes the mixing of flavor eigenstates and mass eigenstates. These matrices indicate that there is a "mismatch" between the particles states when they propagate freely and when they take part in the weak interaction. The source of this mismatch has no explanation in quantum field theory and there is no Lagrangian based description of these phenomena. We take it that this is evidence of our *lack of knowledge* about these systems and about our inability to define the asymptotic states in a coherent way. So given the state of art, it is possible that the weak interaction that produces these oscillating states is non-unitary.

Second, and no less important, the asymptotic oscillating states are *not* superpositions of out-states with different particles content of the *same* interaction, which our proposed collapse postulate excludes. The interaction produces a *distinct* flavor asymptotic state almost instantaneously, but this state propagates as a mass eigenstate which is detected as different flavors, depending on the ratio of the distance from the location of the interaction to the particle's energy. So, this out-state describes a distinct particle that has a unique propagation pattern that oscillates between different flavor states (rather than a superposition of different particles types), each propagating in a unique pattern. It may be that transitions between different flavors of a certain particle should not be considered as particles content change, but as a kind of "polarization" or "spin states" of the *same* fundamental particle. However, this concept cannot be asserted before the source of the oscillation phenomena is better understood.

### 10.3 Formation of bound states

Some interactions which result in bound out-states seem to have a unitary evolution description in the formalism of quantum mechanics, e.g., the Hydrogen atom (electron and proton), Cooper pairs (two electrons), etc. These bound states are the outcomes of interactions where the in-state has a different particles content than that of the out-state, e.g., a free electron and a free proton at the in-state and the Hydrogen atom at the out-state. This seems to contradict our claim that particles content change is the result of a non-unitary process in quantum field theory, which appears as a "collapse of the wave-function" description in quantum mechanics.

However, these quantum mechanical descriptions of bound states formation are *non*-relativistic and are low order approximations of the perturbative quantum field theory description. In this latter description, a bound state is considered a particle associated with an almost local field composed of the constituent fields corresponding to the in-state particles.[30] When these bound states are formed by a weakly coupled theory (e.g., quantum electrodynamics), the coupling "necessarily involve[s] an infinite number of interactions between the constituent particles, and hence an infinite number of Feynman graphs" (Duncan 2012, p. 375). The formation of the Hydrogen stable bound state, in which the constituent particles (the electron and the proton), continue to interact over an infinite time period is an example of such a process. In certain cases, where a *non*-relativistic asymptotic state appears, a sum of finite subset of these graphs can be described by the Bethe–Salpeter wavefunction and provides a practical approximation for the un-summable infinite actual process.

We take it that the unitary descriptions of certain bound states formation are only rough approximations in certain limiting cases of bound states formed in weakly coupled theories. In strongly coupled theories such as quantum chromodynamics, where (for example) the proton is a bound state of three quarks and the nucleus of an atom is a bound state of protons and neutrons, such approximated descriptions are *not* available. It follows that these unitary descriptions of bound states do not present a genuine challenge to our claim that particles content change can only be achieved via a non-unitary process (see Duncan 2012, Ch. 11, for a detailed discussion of bound states formation in quantum field theory).

## 11. Conclusions

We proposed a solution to the measurement problem in quantum mechanics based on the interaction picture in quantum field theory. We suggested that indeed there are two different types of processes in elementary interactions. The distinction between them is based on a well-defined physical criterion of "particles content change" which we elaborated on. There are many natural processes, such as unstable particles decay, particles scattering and absorption and emission of radiation in which this criterion is fulfilled. In the context of the measurement problem, we demonstrated the explanatory power of our proposal by showing that processes defined vaguely as "measurements" invariably include a stage in which such particles content change occurs.

---

[30] Particles obtained from the vacuum by the action of fundamental fields in the first degree are called elementary; the other particles are called composite particles.

Our proposal is supported by Haag's theorem and its extensions. It is a consequence of these theorems that there is *no* unitary evolution from the in-state to the out-state of non-trivial interactions. It follows that assuming a unitary evolution in the interaction picture of quantum field theory is unsound. We further argued that the Haag-Ruelle scattering theory cannot circumvent the implication of Haag's theorem since it cannot describe the gauge invariant interactions of the standard model. As for the corrections to the interaction picture set by renormalization methods and by effective theories, which violate the assumptions of Haag's theorem, we used a different version of this theorem and argued that they also cannot be described by unitary processes.

Our approach is based on existing physical theories. It proposes a trigger for the "collapse of the quantum state which is based solely on the well accepted quantum field theory without introducing an additional external stochastic mechanism. We take this to be an advantage over spontaneous collapse theories, such as the GRW theory.

Since our collapse interpretation proposes that the state of a system in an interaction where there is a particles content change evolves by a non-unitary stochastic transition, there is a question about whether this transition satisfies *Lorentz invariance*, as expected from a relativistic theory. It has been recently argued (see Jones, Guaita and Bassi 2021) that the collapse structure of relativistic extensions of the GRW theory (see e.g., Tumulka 2006, 2020) is *not* Lorentz invariant. Since the nature of the collapse in our proposal is very different from that of the GRW theory, it is not clear whether a similar argument can be made against our proposal. We intend to examine this question in detail in future research.

## 12 Appendices

### 12.1 Locality and the cluster decomposition principle

In this appendix we analyze an interaction between a particle which has an equal probability of being in two separate positions, where two detectors (each represented by a single particle) are located, so that it seems as if the impacting particle can interact simultaneously with both. Let us denote the impacting particle with index $1$ and the two other particles with indices $2, 3$;

and denote the in-state by $\boldsymbol{\alpha} = q_1 q_2 q_3$, and the out-state by $\boldsymbol{\beta} = q'_1 q'_2 q'_3$ where $q_i$ denotes the momentum of the corresponding particle.

The S-matrix of the interaction can be decomposed as follows:
$$S_{\boldsymbol{\beta\alpha}} = S_{\boldsymbol{\beta\alpha}}^C + \delta(q'_1 - q_1) S_{q'_2 q'_3; q_2 q_3}^C + \delta(q'_2 - q_2) S_{q'_1 q'_3; q_1 q_3}^C + \delta(q'_3 - q_3) S_{q'_2 q'_1; q_2 q_1}^C$$
$$+ \delta(q'_1 - q_1) \delta(q'_2 - q_2) \delta(q'_3 - q_3)$$
where the first term is the three particles' connected component, the next three terms are all the possibilities for two particles' connected components (and one particle unaffected), and the last term is the no-interaction term (see Duncan 2012, p. 135).

The first and second terms must nullify since the two particles $2, 3$ are space-like separated and by the cluster decomposition theorem, such interactions cannot occur. Therefore, if an interaction occurs, the probability amplitude $S_{\boldsymbol{\beta\alpha}}$ consists of the third and fourth terms only. For simplicity, consider an ideal interaction for which there exists a frame of reference where $q_1 \neq 0, q_2 = q_3 = q'_1 = 0$ i.e., particles $2, 3$ are stationary before the interaction and particle $1$ after the interaction. Then, in this frame, total momentum conservation implies $q_1 + \cancel{q_3} - \cancel{q'_1} + q'_3 = 0$, $q_2 = q'_2 = 0$ or $q_1 + \cancel{q_2} - \cancel{q'_1} + q'_2 = 0$, $q_3 = q'_3 = 0$ and it is impossible that both are satisfied together. Therefore, only one of these interactions can occur.

A complementary and heuristic argument would be that if a simultaneous detection of particle $1$ is possible, then it would allow superluminal signaling when the detectors are space-like separated. In this hypothetical scenario switching off or turning on one detector influences the probability of detection in the other detector, and hence the no-signaling property is lost. Indeed, if the probability $p_M$ for mutual detections is positive and the probability for a single detection is $p_S$, such that $p_M + 2p_S = 1$, then each detector has a detection probability of $p_M + p_S = 1/2 + p_M/2 > 1/2$. However, in the case that one detector is turned off, the probability of detection in the other detector changes to $1/2$.

These arguments show that in order to satisfy locality in the relativistic sense, or equivalently adhere to no-signaling, a localized position-based collapse must occur in such a measurement scenario.

## 12.2 Nonunitary processes in measurement scenarios

We describe here a few examples of measurement scenarios confirming our conjecture that in every measurement process there is at least *one* stage at which a non-unitary interaction which changes the particles content occurs.

### 12.2.1 Position measurement in the double slit experiment

In the double-slit experiment a beam of light evolves according to the electromagnetic field equation through the two slits and forms an interference pattern on the screen. For a single photon, the probability of hitting the screen in a certain region is given by the amplitude of this interference pattern at that region. When the photon hits the screen its wave description collapses and there is only one position in which it interacts with the screen. This interaction can be described in quantum field theory as an absorption (or scattering) of the photon by a specific atom or molecule of the screen and due to the locality property it can occur only at one position. This absorption/scattering, as explained in Section 8, results in a particles content change, which triggers a "collapse of the wavefunction" description. In this case the position measurement is realized by the macroscopic dot formed on the screen at the location of the interaction.

### 12.2.2 Measurement of momentum in an ionization chamber

A measurement of a particle's momentum in an ionization chamber is computed from the trajectory traces of the particle in the chamber. Regardless of the method of detecting the trajectory, it consists of many consecutive position measurements, and in certain cases also temporal measurements. Hence, the momentum value is in fact computed by integrating over the outcomes of the consecutive measurements. In this case there is a consecutive series of non-unitary interactions (e.g., ionization) of the particle with atoms or molecules in the detector medium. Each one of these interactions results in a collapse that is localized in space-time and the whole series of these locations forms the trajectory.

### 12.2.3 Polarization measurement in the EPR experiment

The experiment (see Aspect 1982) includes three stages. In the first stage the entangled photons go through the polarimeters which split the wavefunction according to the orientation of the polarization angle. This stage is unitary and there is no particles content change. In the second stage the detection occurs when the photon interacts with one of the detectors (or more accurately with a microscopic part of it) and an electron is emitted. In this interaction the photon is (partially or completely) absorbed, the atom is ionized, and an electron is emitted. Hence there is a particles content change and the wavefunction "collapses". In the third stage the emitted electrons are amplified, and a macroscopically observable effect is produced. The polarization of the photon is deduced from the location of the detector in which the photon is detected.

**12.2.4 Measurement of electromagnetic radiation frequency**

Such measurements are performed by spectrum analyzers and there are three major modes of operations: The first one is recording the temporal detection of (low frequency) photons, in a well-localized detector, and then computing the frequency from the full detection series. The second one is the use of an array of detectors, each of which is sensitive to a different narrow frequency range. Here, the measured frequency is deduced from the location in the array of the detector that has actually reacted. The third mode is the use of an apparatus that separates out the light beam into distinct channels, at which detectors that are sensitive to the wavelengths of the radiation are placed, e.g., a prism. Following this separation stage, the signal frequency is determined by identifying the channel whose detector has been triggered by a photon.

In each of these modes, the actual measurement is performed by the detection of a photon in a certain position and time, and then the frequency is computed from this data. In the optical spectrum analyzer case, there is a preparation stage (the splitting of the light beam), which is unitary and does not inflict a change of particles content. The rest of the stages are similar to the stages in the other examples.

**12.2.5 Particle calorimeters**

Particle calorimeters measure the energy of particles. In a typical calorimeter the incoming particle hits a material, called the absorber, and excite the atoms or the atoms' nucleus of this material. As a result of these excitations, the energy of the hitting particle is converted into a shower of particles in the detector, and these particles carry a fraction of the initial particle's energy. Another material, the sampling material, is interleaved within the absorbers, and converts a very small fraction of the shower energy into some measurable quantity: light, electric current, etc. The calorimeter is divided into cells that collect the energy in their vicinity, and the total shower energy is accumulated. The total initial particle energy is computed from the shower energy by estimating which fraction of it stimulates the sampling material. This measurement process is not carried out by position measurements, but rather by localized interactions inside the calorimeter that produce particles and eventually the number of these particles (photons, electrons) is counted in a localized manner. In these measurements there are in fact two non-unitary stages with particles content change. In the first one, interactions with the absorbing medium result in a change of particles content, since new particles are produced. In the second stage, these particles are detected and counted in localized regions of spacetime, and their number is used for computing the original particle energy.

**Acknowledgement & Funding**

This research is supported by the Israel Science Foundation (ISF), grant number 690/21.